\documentclass[a4paper]{jpconf}
\usepackage{amsmath, amsthm, amssymb}
\usepackage{graphicx}
\usepackage{slashed}
\begin{document}
\title{One-loop calculations for $H\rightarrow
f\bar{f}\gamma$ in the $U(1)_{B-L}$ extension 
for Standard Model}
%%%%%%%%%%%%%%%%%%%%%%%%%%%%%%%%%%%%%%%%%%%%%
\author{Khiem Hong Phan$^{(a,b)}$, 
Anh Thu Nguyen$^{(c,d)}$ and 
Dzung Tri Tran$^{(c,d)}$}
\address{
$^{a)}$Institute of Fundamental and Applied 
Sciences, Duy Tan University, Ho Chi Minh 
City $700000$, Vietnam\\
$^{b)}$Faculty of Natural Sciences, 
Duy Tan University, 
Da Nang City $550000$, Vietnam\\
$^{c)}$University of Science, 
Ho Chi Minh City $700000$, Vietnam\\
$^{d)}$Vietnam National University, 
Ho Chi Minh City $700000$, Vietnam}
%%%%%%%%%%%%%%%%%%%%%%%%%%%%%%%%%%%%%%%%%%%%%
\ead{phanhongkhiem@duytan.edu.vn}
%%%%%%%%%%%%%%%%%%%%%%%%%%%%%%%%%%%%%%%%%%%%%
\begin{abstract}
In this paper, we present the calculations
for $H\rightarrow f\bar{f}\gamma$ in the 
$U(1)_{B-L}$ extension for Standard Model. 
Analytic results for one-loop form factors
in the decay process are expressed in 
terms of the scalar one-loop 
Passarino$-$Veltman functions in the 
conventions of {\tt LoopTools}. Therefore, 
the decay rates can be evaluated numerically 
by using this package. 
In phenomenological results,
we show the differential decay rates 
with respect to invariant mass of 
fermion pair
$m_{ff}$, new neutral gauge mass $M_{Z'}$ 
and the coupling $g'$ of $U(1)_{B-L}$ 
gauge group. We find that
the contributions of the 
$U(1)_{B-L}$ extension 
for Standard Model are visible effects 
and they must be taken into account 
at future colliders.
\end{abstract}
%%%%%%%%%%%%%%%%%%%%%%%%%%%%%%%%%%%%%%%%%%%%%
\section{Introduction}%%
%%%%%%%%%%%%%%%%%%%%%%%%
The precise measurements for the 
properties of Standard Model like (SM-like) 
Higgs boson are one of the main 
targets at the High 
Luminosity Large Hadron Collider 
(HL-LHC)~\cite{Liss:2013hbb,CMS:2013xfa}
and future lepton
colliders~\cite{Baer:2013cma}.
It means that all the Higgs 
productions and decay processes 
should be probed as accurately 
as possible. By analyzing 
these high-precision 
experimental data, 
one can verify the Higgs 
sector for answering the nature 
of electroweak symmetry 
breaking (EWSB). It is 
well-known that many of beyond
the Standard Models (BSMs) 
have extended the Higgs sector
as well as gauge sector. 
As a result, many of BSMs 
propose new heavy particles such as
heavy gauge bosons, heavy 
fermions as well as
heavy scalar particles. 
The mentioned particles may 
exchange in the loop
of Feynman diagrams in 
Higgs productions
and decay channels.
Therefore, the accuracy of 
the Higgs production 
cross sections and decay rates
are not only used for testing 
the SM at the high energy regions
but also for constraining new 
parameters in many BSMs.
Among Higgs decay channels, 
the decay processes $H\rightarrow
f\bar{f}\gamma$ have
been considerable
interests
recently at the HL-LHC~\cite{CMS:2015tzs,
CMS:2017dyb, CMS:2018myz,ATLAS:2021wwb}.

The computations for one-loop corrections
to the decay channel $H\rightarrow f\bar{f}\gamma$ 
within the SM framework have reported in~\cite{Abbasabadi:1995rc,
Djouadi:1996ws,Abbasabadi:2000pb,Dicus:2013ycd,
Sun:2013rqa,Passarino:2013nka,Dicus:2013lta,
Kachanovich:2020xyg}. While one-loop 
corrections to $H\rightarrow f\bar{f}\gamma$
in the minimal super-symmetric standard model 
have computed in~\cite{Li:1998rp}. 
In the current paper, we present the calculations
for $H\rightarrow f\bar{f}\gamma$ in the 
$U(1)_{B-L}$ extension for Standard Model. 
Analytic results for one-loop form factors
in this decay process are expressed in terms 
of the scalar one-loop Passarino-Veltman 
functions in the conventions of 
{\tt LoopTools}. Therefore, the decay 
rates can be evaluated numerically by using
this package. 
In phenomenological analyses,
we present the differential decay rates 
with respect the to invariant mass of 
fermion pair $m_{ff}$, new gauge boson mass $M_{Z'}$ and the coupling $g'$ of 
the $U(1)_{B-L}$ gauge group. 
The contributions of the $U(1)_{B-L}$ 
extension 
for Standard Model are visible impacts 
and these should be
taken into account at future colliders.
We stress that the detailed of the
phenomenological analyses for
the decay channel in this model and 
in the two-Higgs-doublet model 
can be found
in~\cite{VanOn:2021myp}. 

The outline of this paper is as follows. 
We review the 
$U(1)_{B-L}$ extension 
for Standard Model in section $2$. 
The evaluations for $H\rightarrow
f\bar{f}\gamma$ in this model are also
performed in this section. We then present
phenomenological results for the decay
process in this section.
Conclusions and outlooks are shown 
in the section $3$.
%%%%%%%%%%%%%%%%%%%%%%%%%%%%%%%%%%%%%%%%%%%%%
\section{Calculation}%%
%%%%%%%%%%%%%%%%%%%%%%%
In this section, we first review the 
$U(1)_{B-L}$ extension 
for Standard Model. We then discuss 
the evaluations
for $H\rightarrow
f\bar{f}\gamma$ in $U(1)_{B-L}$ extension 
for standard model. 
%%%%%%%%%%%%%%%%%%%%%%%%%%%%%%%%%%%%%%%%%%%%%
\subsection{Review of 
the $U(1)_{B-L}$ extension 
for Standard Model} %%
%%%%%%%%%%%%%%%%%%%%%%
Reviewing briefly the 
$U(1)_{B-L}$ extension~\cite{Basso:2008iv}
is presented in this subsection. 
The model is based on 
gauge symmetry $SU(3)_C \otimes SU(2)_L \otimes 
U(1)_Y\otimes U(1)_{B-L}$. In the gauge sector
we add new gauge boson $B_{\mu}'$ 
with the coupling $g'$. After  
the spontaneous electroweak symmetry 
breaking, one has $W^{\pm}$, $Z$
and $Z'$ gauge bosons. In the Higgs sector, 
we include a complex scalar $S$. As a result, 
general scalar potential is given by 
\begin{eqnarray}
V(\Phi, S) &=&  m^2 \Phi^{\dag} \Phi 
+ \lambda_1 (\Phi^{\dag} \Phi)^2 
+\mu^2 |S|^2 
+ \lambda_2 |S|^4 
+ \lambda_3 \Phi^{\dag} \Phi |S|^2.
\end{eqnarray}
Expanding the scalar fields around 
their vacuum as follows:
\begin{eqnarray}
\Phi =
\begin{pmatrix}
\phi^+\\
\dfrac{v_{\phi} +h + i \xi}{\sqrt{2}}
\end{pmatrix}
, \quad S = \dfrac{v_{s} + h' +i \xi'}{\sqrt{2}},
\end{eqnarray}
We can observe the mass spectrum 
of the scalar sector. In detail, 
the Goldstone bosons 
$\phi^{\pm}, \xi$ will give the masses 
of $W^{\pm}$ and $Z$ bosons.
Working in the unitary gauge, the mass 
eigenvalues of neutral Higgs are 
obtained by applying the rotation
\begin{eqnarray}
\begin{pmatrix}
 h_1\\
 h_2
\end{pmatrix}
= 
\begin{pmatrix}
c_{\alpha}& -s_{\alpha}\\
s_{\alpha}&   c_{\alpha}\\
\end{pmatrix}
\begin{pmatrix}
 h\\
 h'
\end{pmatrix}
.
\end{eqnarray}
Where the mixing angle is calculated as
\begin{eqnarray}
s_{2\alpha} = \sin2\alpha = \dfrac{\lambda_3 v_{\phi} v_{s}}
{ \sqrt{(\lambda_1 v_{\phi}^2  - \lambda_2 v_{s}^2)^2 
+ (\lambda_3 v_{\phi} v_{s})^2}}.
\end{eqnarray}
With the help of the above transformation, 
we get the masses of scalar Higgs fields
as follows:
\begin{eqnarray}
 M^2_{h_1} &=& \lambda_1 v_{\phi}^2  + \lambda_2 v_{s}^2 -
 \sqrt{(\lambda_1 v_{\phi}^2  - \lambda_2 v_{s}^2)^2  +
 (\lambda_3 v_{\phi} v_{s})^2},\\ 
 M^2_{h_2} &=& \lambda_1 v_{\phi}^2  + \lambda_2 v_{s}^2 +
 \sqrt{(\lambda_1 v_{\phi}^2  - \lambda_2 v_{s}^2)^2 +
 (\lambda_3 v_{\phi} v_{s})^2}.
\end{eqnarray}
By expanding the kinematic terms in the scalar
sector, we then obtain the masses of gauge bosons
$W, Z$ and $Z'$. 
Especially, from the following 
kinematic terms
\begin{eqnarray}
 \mathcal{L}_{\text{K}} \rightarrow
 (D^{\mu}\Phi)^+D_{\mu}\Phi, \quad (D^{\mu}S)^+D_{\mu}S.
\end{eqnarray}
one has the mass 
of $Z'$ is $M_{Z'} = 2 v_{s}g_1'$.

In the model, we take into account three right-handed neutrinos. Thus, the Yukawa interaction
with including
the right-handed neutrinos is expressed as
\begin{eqnarray}
\mathcal{L}_Y &=&
-y^d_{jk} \bar{q}_{Lj}d_{Rk} \Phi 
-y^u_{jk} \bar{q}_{Lj}u_{Rk} i\sigma_2\Phi^* 
-y^e_{jk} \bar{l}_{Lj}e_{Rk} \Phi 
\nonumber\\
&& -y^{\nu}_{jk} \bar{l}_{Lj}\nu_{Rk} i\sigma_2\Phi^* 
-y^{M}_{jk} \overline{(\nu_R)}_{j}^c\nu_{Rk}\; S + \text{h.c}. 
\end{eqnarray}
for $j,k=1,2,3$. The Majorana 
mass terms for right-handed
neutrinos are corresponding to the
last term in above equation. 
The mass matrix
of neutrinos is given after the EWSB
\begin{eqnarray}
M = 
\begin{pmatrix}
0           & m_D = \dfrac{(y^{\nu})^*}{\sqrt{2}}v_{\phi}
\\
m_D^T       & \sqrt{2}y^{(M)}\; v_{s} 
\\ 
\end{pmatrix}
.
\end{eqnarray}
The diagonalization is obtained by the transformation
\begin{eqnarray}
\text{diag}\Big(-\dfrac{(m^i_D)^2}{M^i}, M^i \Big)= 
\begin{pmatrix}
\cos\alpha_i & -\sin\alpha_i \\
\sin\alpha_i &  \cos\alpha_i \\
\end{pmatrix}
\begin{pmatrix}
 0& m_D^i \\
 m_D^i& M^i \\
\end{pmatrix}
\begin{pmatrix}
\cos\alpha_i & \sin\alpha_i \\
-\sin\alpha_i &  \cos\alpha_i \\
\end{pmatrix}
\end{eqnarray}
for $i=1,2,3$ and $\alpha_i= \text{arcsin}(m_D^i/M^i)$. 

All relevant couplings in the decay 
under consideration are shown. 
In the limit of $\alpha_i \rightarrow 0$, 
all the couplings
are presented in Table~\ref{U1-coupling}
%%%%%%%%%%%%%%%%%%%%%%%%%%%%%%%
\begin{center}
\begin{table}[ht]
\centering
{\begin{tabular}{l@{\hspace{4cm}}l }
\hline \hline
\textbf{Vertices} & \textbf{Couplings}\\
\hline \hline 
$h_{1}(h_2)f\bar{f}$ &
$-i\dfrac{c_{\alpha} 
(s_{\alpha})e\; m_f}{2 M_W s_W}$
\\ \\
$Z'_{\mu} f\bar{f}$  & 
$i Q_f\; g'_1\gamma_{\mu}$
\\  \\
$h_{1}(h_2) W^+_{\mu}W^-_{\nu}$ &
$i \frac{eM_W}{s_W} \;c_{\alpha} (s_{\alpha}) g_{\mu\nu}$
\\  \\ 
$h_{1}(h_2) Z_{\mu}Z_{\nu}$ 
&
$i\frac{eM_W}{c_W^2\; s_W} \;c_{\alpha} (s_{\alpha}) g_{\mu\nu}$
\\   \\
$h_{1}(h_2) Z'_{\mu}Z'_{\nu}$ &
$-i 4g'_1 M_Z' \; s_{\alpha} (-c_{\alpha})g_{\mu\nu}$ \\ 
\hline \hline
\end{tabular}}
\caption{
\label{U1-coupling}
All the couplings involving the decay processes
$H\rightarrow f\bar{f} \gamma$ in the $U(1)_{B-L}$ 
extension of the SM.}
\end{table}
\end{center}
%%%%%%%%%%%%%%%%%%%%%%%%%%%%%%%%%%%%%%%%%%%%%
\subsection{One-loop formulas 
for $H\rightarrow f\bar{f}\gamma$}
%%%%%%%%%%%%%%%%%%%%%%%%%%%%%%%%%%%
General one-loop amplitude 
for $H\rightarrow f\bar{f}\gamma$
can be decomposed 
as follows~\cite{Kachanovich:2020xyg}:
\begin{eqnarray}
 \mathcal{A}_{\text{loop}} &=&\sum\limits_{k=1}^2 
 \Big\{
[q_3^{\mu}q_k^{\nu} - g^{\mu\nu}q_3\cdot q_k] \bar{u}(q_1)
(F_{k,R} \gamma_{\mu}P_R + F_{k,L} \gamma_{\mu}P_L) v(q_2) 
\Big\}\varepsilon^{*}_{\nu}(q_3).
\end{eqnarray}
Where one-loop form factors are 
computed as follows:
\begin{eqnarray}
\label{FLR}
 F_{k,L/R} &=& F_{k,L/R}^{\text{$V_{k}^{0*}$-poles}}
 +  F_{k,L/R}^{\text{Non-$V_{k}^{0*}$} }
\end{eqnarray}
for $k=1,2$. All related kinematic invariant
variables are included: 
$q^2 = q_{12} = (q_1+q_2)^2$, $q_{13} = (q_1+q_3)^2$
and $q_{23} = (q_2+q_3)^2$. Here we also used 
$P_{L/R} = (1\mp \gamma_5)/2$. 

The calculations for this decay process 
are explained as follows. All one-loop Feynman 
diagrams for this process in the 
$U(1)_{B-L}$ extension 
for Standard Model are 
shown in~\cite{VanOn:2021myp}. 
Writing down the amplitudes 
for all Feynman diagrams in this 
channel, one then handles 
Dirac traces and Lorentz 
contractions in $d$ dimensions 
with the help of 
{\tt Package-X}~\cite{Patel:2015tea}. As a result, 
the decay amplitude is then expressed
in terms of tensor one-loop integrals. These 
tensor one-loop integrals can be reduced into 
scalar one-loop Passarino-Veltman 
functions (PV-functions)
~\cite{Denner:2005nn}. The PV-functions 
are written in the form 
of {\tt LoopTools}~\cite{Hahn:1998yk}. 
As a result, we can perform numerical
evaluations for the 
decay rates by using this program.
Analytical results
for one-loop form factors are 
presented in detail in 
\cite{VanOn:2021myp}. The decay rates 
are computed in terms of the one-loop
form factors as 
follows~\cite{Kachanovich:2020xyg}: 
\begin{eqnarray}
\label{decayrate}
 \dfrac{d\Gamma}{d q_{12}q_{13}} 
 = \dfrac{q_{12}}{512 \pi^3 M_H^3}
 \Big[
 q_{13}^2(|F_{1,R}|^2 + |F_{2,R}|^2)
 + q_{23}^2(|F_{1,L}|^2 + |F_{2,L}|^2)
 \Big]. 
\end{eqnarray}
The integration region is 
$0 \leq q_{12} \leq M_H^2$ 
and $0\leq q_{13} \leq M_H^2 - q_{12}$. 
%%%%%%%%%%%%%%%%%%%%%%%%%%%%%%%%%%%%%%%%%%%%%
\subsection{Phenomenological results}%%%
%%%%%%%%%%%%%%%%%%%%%%%%%%%%%%%%%%%%%%%%
We present the phenomenological results
for the decay channel in the 
$U(1)_{B-L}$ extension of the SM. 
We use the following input 
parameters for 
numerical study:
$\alpha=1/137.035999084$, 
$M_Z = 91.1876$ GeV, 
$\Gamma_Z  = 2.4952$ GeV, 
$M_W = 80.379$ GeV, $M_H =125.1$ GeV,
$m_{\tau} = 1.77686$ GeV, $m_t= 172.76$ GeV, 
$m_b= 4.18$ GeV, $m_s = 0.93$ GeV 
and $m_c = 1.27$ GeV. We include 
three more parameters such as
the coupling $g'$, boson mass 
$M_{Z'}$ and the mixing 
angle $\alpha$ 
(or $c_{\alpha}, s_{\alpha}$). 
%%%%%%%%%%%%%%%%%%%%%%%%%%%%%%%%%%%%%%%%%%%%%

\begin{figure}[ht]
\centering
$\begin{array}{cc}
\hspace*{-4.5cm}
d\Gamma/d g' [\textrm{GeV}]
& 
\hspace*{-4.5cm}
d\Gamma/d g' [\textrm{GeV}] \\
\includegraphics[width=7.7cm, height=7.5cm]
{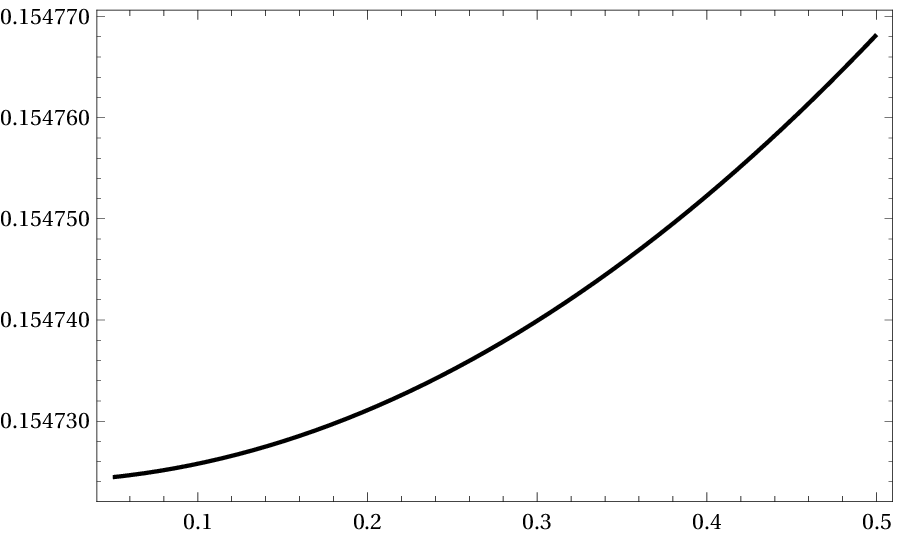}
&
\includegraphics[width=7.7cm, height=7.5cm]
{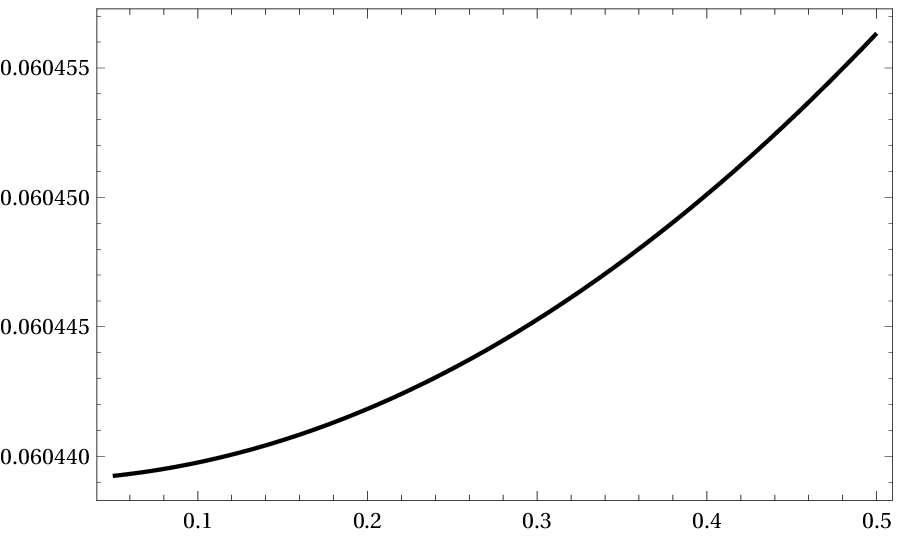}
\\
\hspace{7cm}
g' &
\hspace{7cm}
g'
\end{array}$
\caption{\label{gammagamma}
Differential 
decay rates with respect 
to $g'$ for $M_{Z'}=800$ GeV
(left panel)
and for $M_{Z'}=1000$ GeV 
(right panel).}
\end{figure}

In the Fig.~\ref{gammagamma}, differential 
decay rates with respect to $g'$ for 
$M_{Z'}=800$ GeV (left panel)
and for $M_{Z'}=1000$ GeV (right panel) are shown.
In both Figures, we fix the mixing angle 
$c_{\alpha}=0.5$. 
We find that the decay rates are
proportional 
to $g'$. Furthermore, one can verify that 
decay rates are inversely proportional
to $M_{Z'}$. 
%%%%%%%%%%%%%%%%%%%%%%%%%%%%%%%%%%%%%%

\begin{figure}[ht]
\centering
$\begin{array}{cc}
\hspace*{-4.5cm}
d\Gamma/d m_{ff} [\textrm{GeV}]
& 
\hspace*{-4.5cm}
d\Gamma/d M_{Z'} [\textrm{GeV}] \\
\includegraphics[width=7.7cm, height=7.5cm]
{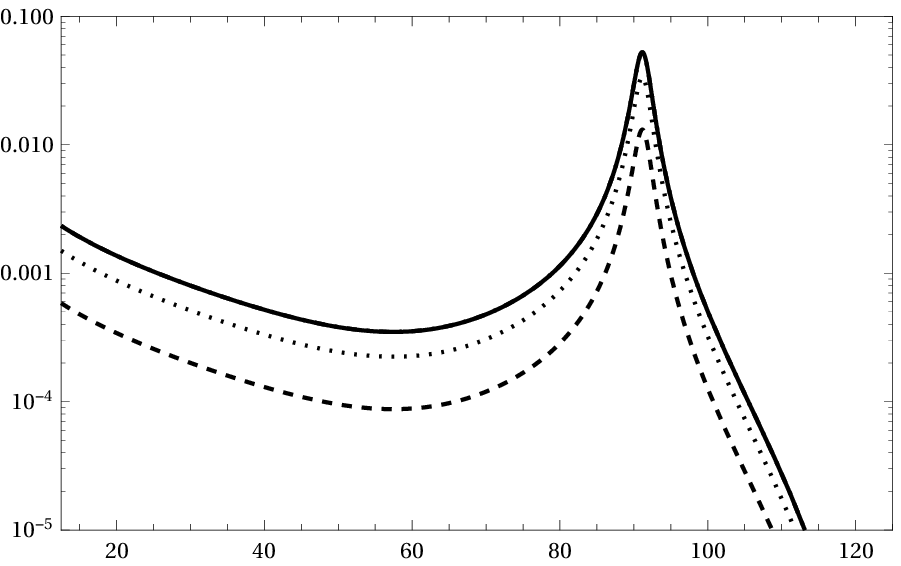}
&
\includegraphics[width=7.7cm, height=7.4cm]
{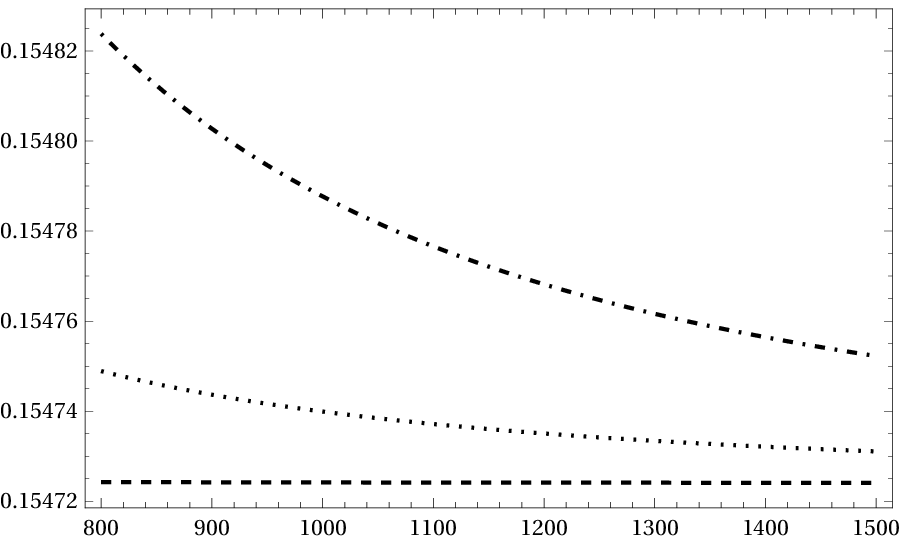}
\\
\hspace{6cm}
m_{ff} [\textrm{GeV}]  &
\hspace{6cm}
m_{Z'} [\textrm{GeV}]
\end{array}$
\caption{\label{Zgamma} Differential 
decay rates with respect to the 
invariant mass of fermion pair 
$m_{ff}$ (left panel)
and $M_{Z'}$ (right panel).}
\end{figure}

In the Fig.~\ref{Zgamma}, 
differential 
decay rates with respect to the 
invariant mass of fermion pair 
$m_{ff}$ (left panel)
and $M_{Z'}$ (right panel) are 
plotted. In the left 
Figure, the solid line shows for the case 
of $c_{\alpha}=1$ 
(it is also corresponding to 
the SM case), 
the dashed line presents for the case of 
$c_{\alpha}=0.8$ and the dotted line is for 
for the case of 
$c_{\alpha}=0.5$, respectively. In the right
Figure, we select the values of $c_{\alpha}=0.3, 
0.5, 0.8$ 
and change $M_{Z'}$ from $800$ GeV to
$1500$ GeV. In this Figure, 
the dashed line presents for the case of 
$c_{\alpha}=0.3$, the dotted line is 
for the case of 
$c_{\alpha}=0.5$ and the dotted 
dashed line
is for the case of 
$c_{\alpha}=0.8$. One observes the peak 
of the decay rates at $m_{ff}=M_Z=
91.1876$ GeV which is corresponding 
to the resonance
of $Z\rightarrow f\bar{f}$.
We find that the decay 
rates are  inversely proportional
to the $M_{Z'}$.  
It is interested to observe that 
the contributions from the 
$U(1)_{B-L}$ extension of the SM
are visible impacts. These effects
must be taken into account at future
colliders. 
%%%%%%%%%%%%%%%%%%%%%%%%%%%%%%%%%%%%%%%%%%%%%

\begin{figure}[ht]
\centering
\includegraphics[width=14cm, height=14cm]
{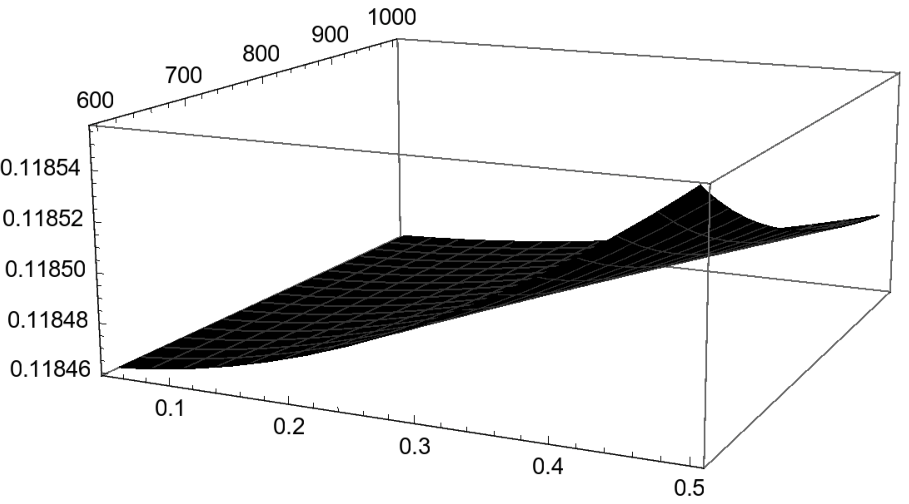}
\caption{\label{3d} Differential 
decay rates with respect to $M_{Z'}$ and $g'$. }
\end{figure}

%%%%%%%%%%%%%%%%%%%%%%%%%%%%%%%%%%%%%%%%%%%%%
In the Fig.~\ref{3d}, differential 
decay rates with respect to $M_{Z'}$ 
and $g'$ are shown. We vary the coupling 
$0.01\le g' \le 0.5$ and boson mass $600$ GeV
$\le M_{Z'}\le 1500$ GeV.
We have the same 
conclusions as previous
cases that the decay 
rates are proportional to 
$g'$ and inversely proportional to 
the $M_{Z'}$. Full analyses for
the decay channel in this model and 
in the two-Higgs-doublet model 
can be found in~\cite{VanOn:2021myp}.
%%%%%%%%%%%%%%%%%%%%%%%%%%%%%%%%%%%%%%%%%%%%%
%%%%%%%%%%%%%%%%%%%%%%%%%%%%%%%%%%%%%%%%%%%%

%%%%%%%%%%%%%%%%%%%%%%%%%%%%%%%%%%%%%%%%%%%%%
\section{Conclusions}
%%%%%%%%%%%%%%%%%%%%%%%%%%%%%%%%%%%%%%%%%%%%%
In the current paper, we have presented the 
calculations
for $H\rightarrow f\bar{f}\gamma$ in the 
$U(1)_{B-L}$ extension for Standard Model. 
Analytic results for one-loop form factors
in this decay process are expressed in terms 
of the scalar one-loop Passarino-Veltman 
functions in the conventions of 
{\tt LoopTools}. Therefore, the decay 
rates can be evaluated numerically by using
this package. 
In phenomenological results,
we have shown differential decay rates 
with respect to $m_{ff}$, $M_{Z'}$ 
and $g'$. We find that
the contributions of the 
$U(1)_{B-L}$ extension for Standard Model 
are visible effects and they must be
taken into account at future colliders.
%%%%%%%%%%%%%%%%%%%%%%%%%%%%%%%%%%%%%%%%%%%%%
\\

{\bf Acknowledgment:}~
This research is funded by Vietnam National
University, Ho Chi Minh City (VNU-HCM) 
under grant number C$2022$-$18$-$14$.\\ 
%%%%%%%%%%%%%%%%%%%%%%%%%%%%%%%%%%%%%%%%%%%%%

%%%%%%%%%%%%%%%%%%%%%%%%%%%%%%%%%%%%%%%%%%%%%
\section*{References}
%%%%%%%%%%%%%%%%%%%%%%%%%%%%%%%%%%%%%%%%%%%%%

%%%%%%%%%%%%%%%%%%%%%%%%%%%%%%%%%%%%%%%%%%%%%
\end{document}